\documentclass[preprint]{aastex}
\usepackage{amsbsy} 

\newcommand{\Lsun}{L_\odot}
\newcommand{\Msun}{M_\odot}

\newcommand{\bu}{\boldsymbol{u}}
\newcommand{\bx}{\boldsymbol{x}}

\newcommand{\bdot}{\boldsymbol{\cdot}}
\newcommand{\bnabla}{\boldsymbol{\nabla}}
\newcommand{\btimes}{\boldsymbol{\times}}

\slugcomment{to appear in MNRAS}

\shorttitle{The Orbital Decay of Embedded Binary Stars}
\shortauthors{Stahler}

\begin{document}

\title{The Orbital Decay of Embedded Binary Stars}

\author{Steven W. Stahler\altaffilmark{1}}

\altaffiltext{1}{Astronomy Department. University of California,
Berkeley, CA 94720}

\email{Sstahler@astro.berkeley.edu}

\begin{abstract}
Young binaries within dense molecular clouds are subject to dynamical friction 
from ambient gas. Consequently, their orbits decay, with both the separation 
and period decreasing in time. A simple analytic expression is derived for this
braking torque. The derivation utilizes the fact that each binary acts as a 
quadrupolar source of acoustic waves. The acoustic disturbance has the 
morphology of a two-armed spiral and carries off angular momentum. From the
expression for the braking torque, the binary orbital evolution is also 
determined analytically. This type of merger may help explain the origin of
high-mass stars. If infrared dark clouds, with peak densities up to  
\hbox{$10^7\,\,{\rm cm}^{-3}$}, contain low-mass binaries, those with
separations less than 100~AU merge within about $10^5$~yr. During the last few
thousand years of the process, the rate of mechanical energy deposition in the 
gas exceeds the stars' radiative luminosity. Successive mergers may lead to the
massive star formation believed to occur in these clouds.
\end{abstract}

\keywords{binaries: general --- stars: formation --- stars: early-type ---
ISM: clouds}

\section{Introduction}

The youngest binaries are still embedded in molecular cloud gas. During this 
phase, the orbiting components experience dynamical friction with the 
surrounding medium. The braking torque causes the stars to spiral inward, so 
that their separation and period diminish with time. Such orbital decay may 
be significant in the very densest molecular clumps. In particular, infrared
dark clouds, currently believed to be the birthsites of massive stars, have
peak number densities as high as \hbox{$10^7~{\rm cm}^{-3}$} \citep{r06,b07}. 
This impressive figure matches that in hot molecular cores, already known to 
contain luminous, high-mass objects \citep{k00}. Binary mergers could be 
occurring in these environments, and could play a role in forming the massive 
stars.

There is still no direct evidence that either infrared dark clouds or hot 
molecular cores contain low-mass stars. However, observations show that, with 
few exceptions, massive stars originate in populous clusters \citep{dw05}. 
Indeed, one account of high-mass star formation is that it proceeds through the
coalescence of lower-mass cluster members \citep{b98,s00}. At even the highest 
observed cluster densities, collisions between stars, either bare or with 
extended disks, occur at too low a frequency. However, the presence of dense, 
background gas should enhance the capture rate \citep{bz05}. Specifically, the 
braking torque from dynamical friction might not only shrink initially wide 
binaries, but cause the component stars to merge. An accelerating succession of
such mergers could then build up massive stars. 

To assess this possibility, the present study calculates the torque on a binary
embedded within an extensive gas cloud. Although the torque is created by
dynamical friction, the usual analysis of that effect is difficult to apply in 
this context. Following \citet{d64}, theorists have determined the frictional 
drag on a single mass traveling through ambient gas in a straight line 
\citep[see also][]{rs71,rs80,o99}. Just as in the stellar dynamical case 
\citep{c43}, the moving mass draws external matter into a trailing wake. It is 
the gravitational tug from this wake that provides the effective drag. The same
essential mechanism is at work in the case of a binary. However, the 
trajectories of both perturbing masses are now closed orbits. Further 
complicating the analysis is the fact that each star draws in matter from the 
wake of its companion. Naive application of the standard dynamical friction
formulae would grossly misestimate the torque.

The new approach introduced here concentrates on the fact that the binary as a 
whole must shed angular momentum to the external medium. The actual mechanism 
is that the orbiting stars create an oscillating gravitational potential that
torques nearby gas. This fluctuating torque generates outgoing acoustic waves, 
which transport angular momentum. By calculating the total angular momentum 
efflux from the binary, the braking torque may be obtained without considering 
the complex star-gas interaction close to the stars themselves.

Section~2 below formulates the problem mathematically and derives the 
governing wave equation. The quadrupolar driving potential created by the 
rotating stars is established in Section~3, while Section~4 discusses the 
physical character of the generated waves. Section~5 presents the central 
result of this investigation. Here the angular momentum transported by the
spiral wave is obtained, and thereby the torque (see eq.~(39)). It is also 
shown that the wave carries off all the mechanical energy released by the 
shrinking binary. The resulting temporal evolution of the system is considered 
in Section~6, along with the efficacy of braking in infrared dark clouds. 
Finally, Section~7 compares the results obtained here with the traditional 
analysis and discusses future extensions of this study.

\section{Formulation of the Problem}

\subsection{Length Scales}

We wish to treat the binary as a perturbing mass embedded in otherwise static, 
uniform gas. If the binary itself recently formed, then gas surrounding it 
would not be perfectly quiescent. Hence, our assumed background is a highly 
idealized representation of a real cloud, or at least that portion of a cloud 
in which we can accurately follow the propagation of acoustic waves from the 
stars. One obvious stipulation is that this region cannot be so close to the 
binary that cloud gas is infalling onto the stars. This requirement sets an 
inner radius of validity for the analysis, which we take to be the sonic point 
in the Bondi accretion problem:
\begin{mathletters}
\begin{eqnarray}
r_{\rm in} \,&\equiv&\, G\,M_{\rm tot}/{2\,c_s^2}  \\
             &=&\, 100\,{\rm AU} \,
             \left({{M_{\rm tot}}\over{1\,\,\Msun}}\right)\,
             \left({c_s\over{2\,\,{\rm km}\,\,{\rm s}^{-1}}}\right)^{-2}\,\,.
\end{eqnarray}
\end{mathletters}
Here, $M_{\rm tot}$ is the total binary mass and $c_s$ the sound speed of the
surrounding gas, assumed to be isothermal. In the numerical evaluation of 
$r_{\rm in}$, we have used for $c_s$ the typical observed velocity dispersion 
of infrared dark clouds \citep{s05}.\footnote{We do not know the physical 
origin of the velocity dispersion, which may represent magnetically mediated 
turbulence, as is believed to be the case in local clouds. We follow the 
conventional route of assuming that this motion is isotropic and contributes 
an effective pressure acting on the cloud as a whole. In more complete studies
of molecular clouds, composite equations of state are employed 
\citep[e.g.][]{mh99}. For the present work, a simpler, isothermal equation of 
state suffices \citep[see also][]{hs07}.} Inside $r_{\rm in}$, where the binary
itself may reside, gas falls freely onto the central stars.

If the binary separation, to be denoted as $a_{\rm tot}$, exceeds $r_{\rm in}$,
then each orbiting star is surrounded by its own zone of infall. In that case,
$a_{\rm tot}$ becomes the appropriate inner boundary. This comparison of 
$a_{\rm tot}$ relative to $r_{\rm in}$ effectively distinguishes two cases: 
``hard'' binaries, in which the relative speed of the component stars is 
supersonic with respect to the gas \hbox{$(a_{\rm tot}\,<\,r_{\rm in})$} and 
``soft'' binaries, for which this speed is subsonic 
\hbox{$(a_{\rm tot}\,>\,r_{\rm in})$}. We shall concentrate on the first case,
since it yields, as we shall see in Section~4, a simplified form of the
acoustic disturbance. 

We also require that our propagation region is not so spatially extended that
the dominant gravitational force is from interior gas rather than the stars. 
Thus, we are confined within another radius $r_{\rm gas}$, where the mass of 
background gas rivals that of the stars. For an ${\rm H}_2$ number density of 
$10^7\,\,{\rm cm}^{-3}$, representating the most compact regions of infrared 
dark clouds, we find that $1\,\,\Msun$ is contained in 
\hbox{$r_{\rm gas} \,=\,1600\,\,{\rm AU}$}. Beyond even this point, theory
indicates, and observations confirm \citep[e.g.][]{bt09}, that the mass density
$\rho$ falls off in response to the self-gravity of the cloud as a whole. This 
outermost radius is the Jeans length
$\lambda_J$,
where
\begin{mathletters}
\begin{eqnarray}
\lambda_J \,&\equiv&\, {c_s\over{\sqrt{G\,\rho}}} \\
             &=&\, 9\times 10^3\,\,{\rm AU}\,
            \left({n\over{10^7\,\,{\rm cm}^{-3}}}\right)^{-1/2}\,
            \left({c_s\over{2\,\,{\rm km}\,\,{\rm s}^{-1}}}\right) \,\,.
\end{eqnarray}
\end{mathletters}
Our wave analysis is thus valid within a somewhat restricted, but well-defined,
dynamic range.

By the same token, we {\it cannot} model, using this method, binaries 
embedded within the dense cores characterizing low-mass star formation 
environments \citep[see][for observations of candidate stellar pairs]{h04}. 
Here, the entire cloud mass is comparable to the stellar $M_{\rm tot}$, and
\hbox{$a_{\rm tot}\,<\,r_{\rm in}\,\sim\,r_{\rm gas}\,\sim\,\lambda_J$}. 
Physically, the dynamics inside such a dense core is dominated by infall onto 
the binary, unless the cloud is stabilized by tension in its internal magnetic 
field. Acoustic waves may propagate outside the cloud, but their source would 
be both the stars and the dense core gas.
 
\subsection{Perturbation Analysis}

We now introduce our central object, the binary. Let the component stars have
masses $M_1$ and $M_2$, whose values are in the ratio $q$, where 
\hbox{$q\,<\,1$}. We assume, for simplicity, that the two are on circular 
orbits about their mutual center of mass, with radii $a_1$ and $a_2$. For a 
given $M_{\rm tot}$, $q$, and  \hbox{$a_{\rm tot}\,\equiv\,a_1\,+\,a_2$}, the 
fixed separation between the stars, we want to ascertain the perturbing effect 
of the binary on the surrounding cloud, taken to have a background density of
$\rho_0$.

In the linear approximation valid for small disturbances, momentum 
conservation reads 
\begin{equation}
{{\partial\bu_1}\over{\partial t}} \,=\, 
-{c_s^2\over\rho_0}\,\bnabla\rho_1 \,-\, \bnabla\Phi_1^\ast \,\,.
\end{equation} 
Here we have denoted the small induced velocity by $\bu_1$, and have used the 
same subscript for perturbations in the density and potential. The superscript 
on $\Phi_1^\ast$ emphasizes that this perturbation is due to the stars alone. 
The equation of mass continuity is, again to linear order,
\begin{equation}
{{\partial\rho_1}\over{\partial t}} \,+\,
\rho_0\,\bnabla\bdot\bu_1 \,=\,0 \,\,.
\end{equation}
In practice, the term $\bnabla\Phi_1^\ast$ in the momentum equation is 
relatively small. However, this term cannot be discarded, as it is the stellar
gravity that ultimately drives the outoging waves.

The next step is to manipulate equations~(3) and (4) to obtain a wave equation
for $\rho_1$. Taking the divergence of equation~(3) yields
\begin{displaymath}
\bnabla\bdot {{\partial\bu_1}\over{\partial t}} \,=\,
-{c_s^2\over\rho_0}\,\nabla^2\rho_1 \,-\, \nabla^2\Phi_1^\ast \,\,.
\end{displaymath} 
But the time derivative of equation~(4) may be cast as
\begin{displaymath}
\bnabla\bdot {{\partial\bu_1}\over{\partial t}} \,=\,
-{1\over\rho_0}\,{{\partial^2\rho_1}\over{\partial t^2}} \,\,
\end{displaymath}
Combining the last two equations yields
\begin{equation}
\nabla^2\rho_1 \,-\, {1\over c_s^2}{{\partial^2\rho_1}\over{\partial t^2}}
\,=\, -{{4\pi G\rho_0}\over c_s^2}\,\rho_1^\ast \,\,,
\end{equation}
where $\rho_1^\ast$ obeys Poisson's equation:
\begin{equation}
\nabla^2\Phi_1^\ast \,=\, 4\,\pi\,G\,\rho_1^\ast \,\,.
\end{equation} 

Equation~(5) is the inhomogeneous wave equation used in all analyses of 
dynamical friction in gases. In previous studies, however, the source
density $\rho_1^\ast$ tracks the straight-line trajectory of the gravitating
mass. After solving the wave equation for $\rho_1$, the retarding force on
that mass is calculated. In our case, the gravitating mass is spatially
confined. It is thus more convenient to idealize $\rho_1^\ast$ as being 
non-zero only at the origin. This singular, ``equivalent density'' for the 
binary is derived in the next section, where we also show how its temporal 
change generates acoustic waves.

\section{Binaries as Acoustic Sources}

\subsection{Potential of the Binary}

We have written the source term of the wave equation~(5) in terms of the 
equivalent density $\rho_1^\ast$. To derive this density, we begin with
$\Phi_1^\ast$, the potential due to the binary. For any spatially compact
source, the potential is conveniently expressed as a sum over multipoles.
Equation~(4.2) of \citet{j62} gives the electrostatic expression; this is
readily altered to the gravitational one: 
\begin{equation}
\Phi_1^\ast (\bx) \,=\, -4\,\pi\,G\,\sum_{l = 0}^\infty\sum_{m=-l}^l
{q_{lm}\over{2l\,+\,1}}\,{{Y_{lm} (\theta,\phi)}\over r^{l+1}} \,\,.
\end{equation}
Here, the multipole moments are found from the source density 
\citep[][eq. (4.3)]{j62}:
\begin{equation}
q_{lm} \,\equiv\, \int\!Y_{lm}^\ast (\theta^\prime,\phi^\prime)\,
\left(r^\prime\right)^l\,\rho (\bx^\prime)\,d^3x^\prime \,\,.
\end{equation}
In these two expressions, primed coordinates refer to source points, and
unprimed ones to the field points where $\Phi_1^\ast$ is measured.

Figure~1 shows the binary lying in the $x$-$y$ plane of our coordinate 
system, with its center of mass situated at the origin. Also indicated are
two of the three coordinates $(r,\theta,\phi)$ of a field point. At time $t$,
mass $M_1$ is a distance \hbox{$r^\prime\,=\,a_1$} from the origin, and has
rotated from the $x$-axis by the azimuthal angle 
\hbox{$\phi^\prime\,=\,\omega\,t$}, where $\omega$ is the binary's angular
speed. The other mass $M_2$ is located at \hbox{$r^\prime\,=\,a_2$} and
\hbox{$\phi^\prime\,=\,\omega\,t\,+\,\pi$}. Both masses have polar angles
\hbox{$\theta^\prime\,=\,\pi/2$}. 

Evaluating the $q_{lm}$ from equation~(8), we find that the monopole
\hbox{$(l\,=\,0)$} term is
\begin{eqnarray}
q_{00} \,&=&\, 
Y_{00}^\ast \left({\pi\over2},\omega t\right)\,M_1 \,+\,
Y_{00}^\ast \left({\pi\over2},\omega t+\pi\right)\,M_2 
\nonumber \\
&=&\, {M_{\rm tot}\over\sqrt{4\pi}} \,\,.
\end{eqnarray}
In contrast, all dipole \hbox{$(l\,=\,1)$} terms vanish. For example,
\begin{eqnarray}
q_{11} \,&=&\,-\sqrt{{3\over{8\pi}}}\,
\left(a_1 M_1\,-\,a_2 M_2\right)\,\, 
{\rm exp}\,(-i\omega t)  \nonumber \\
&=&\, 0 \,\,,
\end{eqnarray}
by definition of the center of mass.

The quadrupole \hbox{($l\,=\,2$)} terms are the most interesting for our 
purpose. For \hbox{$m\,=\,0$}, we have
\begin{equation}
q_{20} \,=\,-{1\over 2}\sqrt{{5\over{4\pi}}}\,\,I \,\,,
\end{equation}
where \hbox{$I\,\equiv\,M_1\,a_1^2 \,+\, M_2\,a_2^2\,$} is the binary's moment 
of inertia about its axis of rotation. For \hbox{$m\,=\,1$},
\begin{equation}
q_{21} \,=\,0 \,\,,
\end{equation}
while for \hbox{$m\,=\,2$},
\begin{equation}
q_{22} \,=\, {1\over 4}\sqrt{{15\over{2\pi}}}\,I\,\,
{\rm exp}\,(-2i\omega t) \,\,.
\end{equation}
To complete the set, equation~(4.9) of \citet{j62} tells us that
\begin{eqnarray}
q_{2,-1} \,&=&\, -q_{21}^\ast \nonumber \\
&=&\, 0 \,\,,
\end{eqnarray}
and
\begin{eqnarray}
q_{2,-2} \,&=&\, q_{22}^\ast \nonumber \\
&=&\, {1\over 4}\sqrt{{15\over{2\pi}}}\,I\,\,
{\rm exp}\,(+2i\omega t) \,\,.
\end{eqnarray}

Substitution of these coefficients into equation~(7) yields the potential
through quadrupole order:
\begin{eqnarray}
\Phi_1^\ast \,&=&\, - {{G\,M_{\rm tot}}\over r} \,+\,
{{G\,I}\over{2\,r^3}} \,\left( {3\over 2}\,
{\rm cos}^2\,\theta\,-\,{1\over 2}\right) \nonumber \\
&{\phantom =}&  -{{3\,G\,I}\over{8\,r^3}}\,\,{\rm sin}^2\,\theta\,
\left\{ {\rm exp}\, \left[ +2i(\omega t\,-\,\phi)\right] \,\,+\,\,
{\rm exp}\, \left[ -2i (\omega t\,-\,\phi) \right] \right\} \nonumber
\,\,,
\end{eqnarray}
which simplifies to
\begin{equation}
\Phi_1^\ast \,=\, - {{G M_{\rm tot}}\over r} \,-\,
{{G\,I}\over{2\,r^3}}
\left[{3\over 2}\,{\rm sin}^2\,\theta\,\,
{\rm cos}\,2 (\omega t \,-\,\phi) \,-\,
\left( {3\over 2} {\rm cos}^2\,\theta \,-\,{1\over 2} \right)\right]
\end{equation}
Octupole and higher-order terms also exist, but they are of smaller magnitude.

\subsection{Equivalent Density} 

We now investigate what distribution of matter generates the potential 
$\Phi_1^\ast$. This equivalent density $\rho_1^\ast$ is a simplified
representation of the binary that gives rise, through its fluctuating
potential, to the same acoustic waves. Direct calculation reveals that 
\hbox{$\nabla^2\,\Phi_1^\ast \,=\,0$} outside the origin. Thus, the 
equivalent density is confined to \hbox{$r\,=\,0$}. We already know that 
the monopole term in $\Phi_1^\ast$ is generated by a mass $M_{\rm tot}$ at the 
origin. We shall denote this portion of the equivalent density as 
$\rho_{10}^\ast$, and caution the reader not to confuse the double subscript 
with that in the quadrupole moments $q_{lm}$. We then have  
\begin{equation}
\rho_{10}^\ast \,=\, M_{\rm tot} \,\delta (x^\prime)\,\delta (y^\prime)\,
                     \delta (z^\prime) \,\,.
\end{equation}  
Higher multipole terms in the potential similarly have sources, i.e., further
contributions to the equivalent density, that are combinations of $\delta$ 
functions and their derivatives, as discussed in \citet[][p. 1278]{mf53}.

The quadrupole portion of the equivalent density is illustrated in Figure~2.
Here we show a symmetric placement of point masses that generates the 
corresponding part of $\Phi_1^\ast$. We replace the real binary by two equal 
masses lying along the binary's azimuthal direction within the $x$-$y$ plane. 
Each point is at the same small distance $\epsilon$ from the $z$-axis, and the 
value of each individual mass is $I/2\epsilon^2$. In addition, we place a third
point source, with negative mass $-I/\epsilon^2$, at the origin itself. 

This configuration clearly has no monopole moment $q_{00}$, since its total
mass vanishes. It may also be verified that all dipole moments are zero. The
quadrupole moment for \hbox{$m\,=\,0$} is
\begin{eqnarray}
q_{20} \,&=&\, Y_{20}^\ast \left({\pi\over 2},\omega t\right)\,
\epsilon^2\,\left({I\over{2\,\epsilon^2}}\right) \,+\,
Y_{20}^\ast \left({\pi\over 2},\omega t + {\pi\over 2}\right)\,
\epsilon^2\,\left({I\over{2\,\epsilon^2}}\right) \nonumber \\
&=& -{1\over 2}\,\sqrt{{5\over{4\pi}}}\,\,I \,\,.
\end{eqnarray} 
in agreement with equation~(11). The other quadrupole moments similarly match.

Referring to Figure~2, the corresponding portion of the equivalent density, 
which we denote as $\rho_{12}^\ast$, is
\begin{eqnarray}
\rho_{12}^\ast \,&=&\,
{I\over{2\,\epsilon^2}}\,\,[
\delta (x^\prime -\epsilon\,{\rm cos}\,\omega t)\,\,
\delta (y^\prime - \epsilon\,{\rm sin}\,\omega t)\,\,\delta (z^\prime)
\nonumber\\
&{\phantom =}&\ \ \ \  +\,\,
\delta (x^\prime +\epsilon\,{\rm cos}\,\omega t)\,\,
\delta (y^\prime + \epsilon\,{\rm sin}\,\omega t)\,\,\delta (z^\prime)
\nonumber\\
&{\phantom =}&\ \ \ \  -\,\,2\,\delta (x^\prime)\,\,\delta (y^\prime)\,\,
\delta (z^\prime)]
\,\,.   
\end{eqnarray}
We now expand the arguments of the $\delta$ functions to second order in 
$\epsilon$. For example, we write
\begin{displaymath}
\delta (x^\prime - \epsilon\,{\rm cos}\,\omega t) \,=\,
\delta (x^\prime) \,-\, \epsilon\,{\rm cos}\,\omega t\,\,
\delta^\prime (x^\prime) \,+\,
{\epsilon^2\over 2}\,{\rm cos}^2\,\omega t\,\,
\delta^{\prime\prime} (x^\prime) \,\,. 
\end{displaymath} 
We substitute these series back into equation~(19) and multiply them out,
retaining only terms through order $\epsilon^2$. Many terms cancel, leaving 
\begin{equation}
\rho_{12}^\ast \,=\, \rho_A^\ast \,+\, \rho_B^\ast \,+\, \rho_C^\ast \,\,,
\end{equation}
where
\begin{mathletters}
\begin{eqnarray}
\rho_A^\ast \,&\equiv&\, {I\over 2}\,\,{\rm cos}^2 \omega t\,\,
\delta^{\prime\prime} (x)\,\,\delta (y) \,\,\delta (z)
\\
\rho_B^\ast \,&\equiv&\, {I\over 2}\,\,{\rm sin}^2 \omega t\,\,
\delta (x)\,\,\delta^{\prime\prime} (y) \,\,\delta (z)
\\
\rho_C^\ast \,&\equiv&\, {I\over 2}\,\,{\rm sin}\,2\,\omega t\,\,
\delta^\prime (x)\,\,\delta^\prime (y) \,\,\delta (z) \,\,.
\end{eqnarray}
\end{mathletters}

Notice that $\rho_{12}^\ast$ varies temporally with a frequency twice that of
the binary itself. The reason for this doubling is that the simplified
configuration in Figure~2 repeats itself every half period. That is, any
{\it difference} in the component masses manifests itself only in higher-order
multipole terms. 

The full equivalent density is
\begin{equation}
\rho_1^\ast \,=\,\rho_{10}^\ast \,+\, \rho_{12}^\ast \,\,. 
\end{equation}
Since we have purposefully designed $\rho_1^\ast$ to have the same monopole 
and quadrupole moments as the real binary, the generated potential must be 
given by equation~(16). Nevertheless, an important check is to verify this
fact directly, through integraton of Poisson's equation. We carry out the 
relevant calculation in the Appendix. 

\section{Character of the Wave}

\subsection{Prologue: Static Solution}

To find $\rho_1$, we substitute our expression for $\rho_1^\ast$ into the
wave equation~(5). The static portion of $\rho_1^\ast$, given by 
$\rho_{10}^\ast$ in equation~(17), similarly generates a static contribution to
$\rho_1$, which we denote as $\rho_1^s$. That is, $\rho_1^s$ obeys
\begin{equation}
\nabla^2 \rho_1^s \,=\,-{{4\,\pi\,G\,\rho_0\,M_{\rm tot}}\over c_s^2}\,
\delta (x)\,\,\delta (y)\,\,\delta (z) \,\,.
\end{equation}
Because of the linearity of the full wave equation, $\rho_1^s$ can (and should)
be added to the time-varying part we shall derive shortly.

Equation~(23) is Poisson's equation, describing the potential from a point mass
located at the origin. The value of this fictitious ``mass'' is 
\hbox{$-\rho_0\,M_{\rm tot} /c_s^2$}. We may immediately write down the 
solution:
\begin{equation}
\rho_1^s \,=\, {{\rho_0\,G\,M_{\rm tot}}\over{c_s^2\,r}} \,\,.
\end{equation}
A similar, but modified, expression applies to the wake created by a moving
mass in the traditional dynamical friction problem. \citet{o99} has noted that
equation~(24) is the linear approximation to the full density in a hydrostatic
envelope surrounding a gravitating mass $M_{\rm tot}$:
\begin{equation}
\rho (r) \,=\, \rho_0\,\,{\rm exp}
\left({{G\,M_{\rm tot}}\over{c_s^2\,r}}\right) \,\,.
\end{equation}
The sharp density rise interior to \hbox{$r\,=\,r_{\rm in}$} predicted by 
equation~(25) does not actually occur. As we have described, gas in this region
instead goes into free-fall collapse onto the stars.
 
The linear result given by equation~(24) is thus the appropriate form for the
static density enhancement.\footnote{Actually, equation~(24) only gives the
static density perturbation to leading order. As we shall see in Section~4.3,
another term that dies off as $r^{-3}$ is needed to balance the static part of
the quadrupole potential.} However, $\rho_1^s$ makes no contribution to the
angular momentum transport. In calculating the latter, we shall be multiplying
the density by the induced, azimuthal velocity. Since the latter oscillates
sinusoidally, the product vanishes over a period. We therefore turn to the
oscillating density perturbation.

\subsection{Wave Density}

From now on, we may omit $\rho_{10}^\ast$ when considering the equivalent 
density. The time-varying density perturbation, which we shall continue to
denote simply as $\rho_1$, obeys 
\begin{equation}
\nabla^2 \rho_1 \,-\, {1\over c_s^2}\,{{\partial^2\rho_1}\over{\partial t^2}}
\,=\, -{{4\,\pi\,G\,\rho_0}\over{c_s^2}}\,\rho_{\rm 12}^\ast \,\,,
\end{equation}
where $\rho_{12}^\ast$ is given by equations~(20)-(21). We proceed by finding 
those parts of $\rho_1$ (denoted $\rho_A$, etc.) generated by each additive 
component of $\rho_{12}^\ast$. Linearity of the wave equation ensures that we 
can add these individual solutions to obtain the full one.

Consider first the functions ${\cal D}_A$, ${\cal D}_B$, and ${\cal D}_C$
obeying
\begin{mathletters}
\begin{eqnarray}
\nabla^2 {\cal D}_A \,-\, {1\over c_s^2}\,
{{\partial^2{\cal D}_A}\over{\partial t^2}}
\,&=&\, -{{2\,\pi\,G\,\rho_0\,I}\over{c_s^2}}\,
{\rm cos}^2 \omega t\,\,
\delta (x)\,\,\delta (y)\,\,\delta (z) \\
\nabla^2 {\cal D}_B \,-\, {1\over c_s^2}\,
{{\partial^2{\cal D}_B}\over{\partial t^2}}
\,&=&\, -{{2\,\pi\,G\,\rho_0\,I}\over{c_s^2}}\,
{\rm sin}^2 \omega t\,\,
\delta (x)\,\,\delta (y)\,\,\delta (z) \\
\nabla^2 {\cal D}_C \,-\, {1\over c_s^2}\,
{{\partial^2{\cal D}_C}\over{\partial t^2}}
\,&=&\, -{{2\,\pi\,G\,\rho_0\,I}\over{c_s^2}}\,
{\rm sin}\,2\,\omega t\,\,
\delta (x)\,\,\delta (y)\,\,\delta (z) \,\,.
\end{eqnarray}
\end{mathletters}
If we can find these three functions, then differentiation of their governing
wave equations reveals that 
\begin{mathletters}
\begin{eqnarray}
\rho_A \,&=&\,{{\partial^2 {\cal D}_A}\over{\partial x^2}} \\
\rho_B \,&=&\,{{\partial^2 {\cal D}_B}\over{\partial y^2}} \\
\rho_C \,&=&\,{{\partial^2 {\cal D}_C}\over{\partial x\,\partial y}} \,\,.
\end{eqnarray}
\end{mathletters}

Each of the wave equations (27a)-(27c) may be solved using the retarded Greens
function. Applying equations~(6.54) and (6.66) of \citet{j62}, and integrating
over the $\delta$ functions, we find
\begin{mathletters}
\begin{eqnarray}
{\cal D}_A \,&=&\, {{G\,\rho_0\,I}\over{2\,c_s^2\,r}}\,\,
{\rm cos}^2\,(\omega t \,-\, kr) \\
{\cal D}_B \,&=&\, {{G\,\rho_0\,I}\over{2\,c_s^2\,r}}\,\,
{\rm sin}^2\,(\omega t \,-\, kr) \\
{\cal D}_C \,&=&\, {{G\,\rho_0\,I}\over{2\,c_s^2\,r}}\,\,
{\rm sin}\,2\,(\omega t \,-\, kr) \,\,,
\end{eqnarray}
\end{mathletters}
where the wave number \hbox{$k\,\equiv\,\omega/c_s$}.

In taking spatial derivatives of these last three expressions, we utilize the 
fact that we are in the far-field limit \hbox{($k r\,\gg\,1$)}. To see this,
note first that
\begin{eqnarray}
k^2\,a_{\rm tot}^2 \,&=&\, {{G\,M_{\rm tot}}\over{c_s^2\,a_{\rm tot}}} 
\nonumber \\
&=&\, {{2\,\,r_{\rm in}} \over a_{\rm tot}} \nonumber \,\,.
\end{eqnarray}
Since our field point is located well outside $r_{\rm in}$, we have, for
hard binaries,
\begin{displaymath}
k^2\,r^2 \,\gg\, k^2\,r_{\rm in}^2 \,>\, k^2\,a_{\rm tot}^2 \,\,.
\end{displaymath}
We conclude that
\begin{displaymath}
k^2\,r^2 \,\gg\, 
{{2\,\,r_{\rm in}} \over a_{\rm tot}}
\,>\, 1 \,\,.
\end{displaymath}
Following the usual practice in acoustics \citep[e.g.][Chapter 1]{L78}, we 
apply spatial derivatives only to the phase \hbox{($\omega t \,-\, kr$)}.
Derivatives of the prefactors of ${\cal D}_A$, etc. involving $r$ are smaller
by one or two powers of ${(k\,r)}^{-1}$.

Differentiation of equations~(29a)-(29c), under the far-field approximation,
yields expressions for the density components:
\begin{mathletters}
\begin{eqnarray}
\rho_A \,&=&\, -{{G\,\rho_0\,I\,k^2}\over{c_s^2\,r}}\,{\rm sin}^2\theta\,\, 
{\rm cos}^2\phi\,\,{\rm cos}\,2\,(\omega t\,-\,kr)  \\
\rho_B \,&=&\, +{{G\,\rho_0\,I\,k^2}\over{c_s^2\,r}}\,{\rm sin}^2\theta\,\,
{\rm sin}^2\phi\,\,{\rm cos}\,2\,(\omega t\,-\,kr)  \\
\rho_C \,&=&\, -{{G\,\rho_0\,I\,k^2}\over{c_s^2\,r}}\,{\rm sin}^2\theta\,\,
{\rm sin}\,2\phi\,\,{\rm sin}\,2\,(\omega t\,-\,kr) \,\,.
\end{eqnarray}
\end{mathletters}
Adding these gives the full density perturbation:
\begin{equation}
\rho_1 \,=\, -{{G\,\rho_0\,I\,k^2}\over{c_s^2\,r}}\,\,{\rm sin}^2\theta\,\,
{\rm cos}\,2(\omega t \,-\, kr \,-\,\phi) \,\,. 
\end{equation}

It is important to understand, in a qualitative sense, the amplitude of the
density perturbation in equation~(31). The relative perturbation,
$\rho_1/\rho_0$, created by a simple point mass (a monopole) is of order 
$r_{\rm in}/r$, according to equation~(24). However, our oscillating density 
perturbation is quadrupolar. Thus, the monopole result must be multiplied by 
two powers of $k\,a_{\rm tot}$. The amplitude in equation~(31) is indeed of 
order \hbox{$(r_{\rm in}/r)\,(k\,a_{\rm tot})^2$}.

As expected, the perturbation is an acoustic wave that travels radially 
outward with phase velocity \hbox{$\omega/k\,=\,c_s$}. At any time, the phase 
of the wave is also dependent on $\phi$. In fact, equation~(31) reveals that 
the disturbance may also be viewed as a trailing, two-armed spiral wave, with 
a latitude-dependent amplitude that peaks at the equator 
\hbox{($\theta\,=\,\pi/2$)}. Since \hbox{$kr\,\gg\,1$}, the spiral is tightly 
wrapped, with a relatively small pitch angle.

Figure~3 illustrates the basic geometry of the wave. Shown are wavefronts
(surfaces of constant phase) for the two spiral arms in the equatorial plane.
If we trace one arm around the circle, the radius of the front increases by
\hbox{$\lambda\,\equiv\,2\pi/k$}. However, because a second arm is interleaved,
the actual radial wavelength of the disturbance is $\lambda/2$, with an
associated wavenumber of $2\,k$. The perturbation's angular frequency is
$2\,\omega$, so the outward velocity is again \hbox{$2\omega/2k\,=\,c_s$}.

\subsection{Induced Velocity}

We next determine the velocity created in the gas by the passing wave. Taking
the curl of the momentum equation~(3), we find that
\begin{displaymath}
{{\partial{\phantom t}}\over{\partial t}}
\left(\bnabla\,\btimes\,{\bu}_1\right) \,=\, 0 \,\,.
\end{displaymath} 
Thus, the induced vorticity is independent of time, and is zero for oscillatory
motion. It follows that the velocity may be written as
\begin{equation}
{\bu}_1 \,=\, \bnabla \psi_1 \,\,,
\end{equation}
where $\psi_1$ is the velocity potential. From the mass continuity 
equation~(4), $\psi_1$ obeys
\begin{equation}
\nabla^2 \psi_1 \,=\, -{1\over\rho_0}\,{{\partial\rho_1}\over{\partial t}}
\,\,.
\end{equation}

If we assume that $\psi_1$ depends on the same phase as $\rho_1$, then the
dominant contribution to $\nabla^2\psi_1$ in the far-field limit is simply
$-4\,k^2\,\psi_1$. Using $\rho_1 (t)$ from equation~(31), we find that
\begin{equation}
\psi_1 \,=\, 
{{G\,I\,\omega}\over{2\,c_s^2\,r}}\,\,{\rm sin}^2\theta\,\,
{\rm sin}\,2(\omega t \,-\, kr \,-\,\phi) \,\,.
\end{equation}
Finally, we may read off from equation~(32) the velocity components: 
\begin{mathletters}
\begin{eqnarray}
u_r \,&=&\,-{{G\,I\,\omega^2}\over{c_s^3\,r}}\,\,{\rm sin}^2\theta\,\,
{\rm cos}\,2(\omega t \,-\, kr \,-\,\phi) \\
u_\theta \,&=&\,+{{G\,I\,\omega}\over{2\,c_s^2\,r^2}}\,\,{\rm sin}\,2\theta\,\,
{\rm sin}\,2(\omega t \,-\, kr \,-\,\phi) \\
u_\phi \,&=&\,-{{G\,I\,\omega}\over{c_s^2\,r^2}}\,\,{\rm sin}\,\theta\,\, 
{\rm cos}\,2(\omega t \,-\, kr \,-\,\phi) \,\,.
\end{eqnarray}
\end{mathletters}

In deriving $u_r$, we again applied the radial derivative to the phase only. We
see also that this velocity component is larger than the other two by a factor
of order \hbox{$\omega\,r/c_s\,=\,k\,r \,\gg\,1$}. Such dominance of the radial
velocity is expected for a wavefront with small pitch angle. While relatively
small, the $\phi$-component is critical for angular momentum transport.

We may also obtain ${\bu}_1$ directly from the momentum equation~(3). The 
gradient of the monopole contribution to $\Phi_1^\ast$ is balanced by the
static density perturbation previously derived. Equation~(16) shows that there
is also a static part of the quadrupolar potential. This is balanced by a
smaller term in the static density perturbation. That is, the total static 
perturbation, to quadrupole order, is
\begin{equation}
\rho_1^s \,=\, {{\rho_0\,G\,M_{\rm tot}}\over{c_s^2\,r}} \,-\,
{{\rho_0\,G\,I}\over{2\,c_s^2\,r^3}} \,
\left({3\over 2}\,{\rm cos}^2\,\theta\,-\,{1\over 2}\right) \,\,.
\end{equation}

The remaining, oscillatory part of $\Phi_1^\ast$ contributes in principle to
the fluctuating velocity. However, if we actually compare its gradient to the
force associated with the pressure gradient, we find the $\Phi_1^\ast$-gradient
to be smaller by several powers of \hbox{$(k\,r)^{-1}$}. In the far-field 
limit, therefore, the velocity is actually generated only by $\rho_1$, as 
equation~(33) already indicates. If we express the oscillating part of 
$\rho_1$ as the real part of a complex exponential, and set
\begin{displaymath}
{{\partial{\bu}_1}\over{\partial t}} \,=\,2\,i\,\omega\,{\bu}_1 \,\,,
\end{displaymath} 
then we may solve the momentum equation for ${\bu}_1$ itself, obtaining the 
same velocity components as above.
   
\section{Angular Momentum and Energy Transport}

\subsection{Braking Torque}

Imagine surrounding the binary with a spherical shell of radius $r$. We wish to
determine the outflow of angular momentum through this shell. For the acoustic
wave to have the properties we ascribed to it, $r$ must lie between 
$r_{\rm in}$ and $r_{\rm gas}$. The $z$-component of specific angular momentum 
at any point on the shell is \hbox{$r\,{\rm sin}\,\theta\,u_\phi$}, with 
$u_\phi$ given by equation~(35c). Additionally, the mass flux through this 
same point is $\rho\,u_r$, where $u_r$ is taken from equation~(35a). The flux 
of angular momentum, which we denote as $j$, is therefore
\begin{mathletters}
\begin{eqnarray}
j \,&=&\, \rho\,r\,{\rm sin}\,\theta\,\,u_r\,u_\phi \\
    &=&\, {\omega^3\over c_s^5}\,{{\rho_0\,G^2\,I^2}\over r^2}\,
       {\rm sin}^4\theta\,\,{\rm cos}^2\,2(\omega t \,-\,kr\,-\,\phi) \,\,.
\end{eqnarray}
\end{mathletters}
In the last expression, we have used for $\rho$ its equilibrium value $\rho_0$.
Since the product \hbox{$r\,{\rm sin}\,\theta\,u_r\,u_\phi$} already falls off
as $r^{-2}$, any density variation that declines with radius does not appear
in the total angular momentum efflux, integrated over the sphere.

By angular momentum conservation, this outflow, which we denote $\dot J$, must
also be $-\Gamma$, where $\Gamma$ is the torque exerted on the binary by 
surrounding gas. That is
\begin{equation}
\Gamma \,=\, -r^2 \int_0^\pi\!d\theta\,{\rm sin}\,\theta 
\int_0^{2\pi}\!d\phi\,\,j \,\,.
\end{equation}
Using
\begin{displaymath}
\int_0^\pi\!d\theta\,\,
{\rm sin}^5\theta \int_0^{2\pi}\!d\phi\,\,
{\rm cos}^2\,2(\omega t \,-\,kr\,-\,\phi) \,=\,{{16\,\pi}\over 15}\,\,,
\end{displaymath}
we arrive at our main result:
\begin{equation}
\Gamma \,=\, -{{16\,\pi}\over 15}\,{\omega^3 \over c_s^5}\,\rho_o\,G^2\,I^2 
\,\,.
\end{equation}

A striking aspect of the torque is its high sensitivity to the sound speed 
$c_s$. It is more difficult to gather hotter gas into the wakes that actually 
provide the gravitational tug on the orbiting stars. An inverse dependence on 
$c_s$ is also present in the expressions for $\rho_1$ and ${\bu}_i$ (see 
equations~(31) and (35a)-(35c)). In any event, the sensitivity of $\Gamma$ to 
$c_s$ means that quantitative conclusions regarding astrophysical effects of 
the torque are necessarily rather imprecise. 

\subsection{Energy Loss}

The outgoing acoustic wave transports not only angular momentum, but also
mechanical energy. We first note, from equations~(35a)-(35c) and the succeeding
comments, that the kinetic energy density in the far field simplifies to
\begin{displaymath}
{1\over 2}\,\rho\,(u_r^2\,+\,u_\theta^2\,+\,u_\phi^2) \,\rightarrow \,
{1\over 2}\,\rho_0\,u_r^2 \,\,.
\end{displaymath}
Now the {\it total} energy density of any acoustic wave, including the 
component associated with compression by the enhanced pressure, is twice this
kinetic value \citep[][Section~1.3]{L78}. Since the wave travels radially
outward at the sound speed, the total energy flux past any point is
\begin{equation}
{\dot{\cal E}} \,=\, \rho_0\,u_r^2\,c_s \,\,.
\end{equation}

It is instructive to compare this result with $j$, the angular momentum flux in
equation~(37a). From equations~(35a) and (35c), we have
\begin{displaymath}
u_\phi \,=\,{c_s\over\omega}\,
{u_r\over{r\,{\rm sin}\,\phi}} \,\,.
\end{displaymath}
Thus, the angular momentum flux may be written as
\begin{equation}
j \,=\,{c_s\over\omega}\,\rho_0\,u_r^2 \,\,, 
\end{equation}
where we have again replaced the density by its equilibrium value. Integrating
${\dot{\cal E}}$ and $j$ over the entire shell, we obtain a relationship 
between the global energy loss rate $\dot E$ and $\dot J$:
\begin{equation}
{\dot E} \,=\, \omega\, {\dot J} \,\,.
\end{equation}

Since both the energy and angular momentum are being extracted from the binary,
the same relationship between their loss rates should apply to that system. We
now show that this is the case. Referring back to Figure~1, the angular 
momentum of the binary is
\begin{eqnarray}
J_{\rm bin} \,&=&\, M_1\,a_1^2\,\omega \,+\, M_2\,a_2^2\,\omega  \nonumber \\
   &=& I\,\omega \,\,.
\end{eqnarray}
The binary's total energy is
\begin{displaymath}
E_{\rm bin} \,=\, {1\over 2}\,I\,\omega^2 \,-\, 
{{G\,M_1\,M_2}\over a_{\rm tot}} \,\,.
\end{displaymath}
But we also have
\begin{equation}
{{G\,M_{\rm tot}}\over a_{\rm tot}^3} \,=\, \omega^2 \,\,.
\end{equation}
From this equation, applying standard manipulations, we find that the 
potential energy is
\begin{displaymath}
-{{G\,M_1\,M_2}\over a_{\rm tot}} \,=\, -I\,\omega^2 \,\,,
\end{displaymath}
so that
\begin{equation}
E_{\rm bin} \,=\, -{1\over 2}\,I\,\omega^2 \,\,. 
\end{equation}
Comparison with equation~(43) reveals that
\begin{equation}
E_{\rm bin} \,=\, -{\omega\over 2}\,J_{\rm bin} \,\,.
\end{equation}

We next relate the temporal change of $J_{\rm bin}$ to that of $\omega$. We 
first note that the binary's moment of inertia may be written in terms of the
separation $a_{\rm tot}$ and mass ratio $q$:
\begin{equation}
I \,=\, {q \over (1\,+\,q)^2}\,\,M_{\rm tot}\,\,a_{\rm tot}^2 \,\,.
\end{equation}
Thus, if we again use equation~(44) to eliminate $a_{\rm tot}$, the angular 
momentum may be written as
\begin{equation}
J_{\rm bin} \,=\,{q \over (1\,+\,q)^2}\,\,G^{2/3}\,\,M_{\rm tot}^{5/3}\,\,
\omega^{-1/3} \,\,.
\end{equation}
During contraction of the binary, therefore,
\begin{equation}
{{{\dot J}_{\rm bin}}\over J_{\rm bin}} \,=\, -{1\over 3}\, 
{{\dot\omega}\over\omega} \,\,.
\end{equation}
Taking the time derivative of equation~(46) and applying equation~(49) now 
gives
\begin{eqnarray}
{\dot E}_{\rm bin} \,&=&\, 
-{{\dot\omega}\over 2}\,J_{\rm bin} \,-\,
{\omega\over 2}\,{\dot J}_{\rm bin} \nonumber \\
&=&\, {{3\,\omega}\over 2}\,{\dot J}_{\rm bin} \,-\, 
{\omega\over 2}\,{\dot J}_{\rm bin} \nonumber \\
&=&\, \omega\,{\dot J}_{\rm bin} \,\,. 
\end{eqnarray}
As claimed earlier, the energy and angular momentum of the binary change at the
same relative rates as these same quantities in the outgoing wave.

The rate of energy transport by the wave can be recast in another way that
provides a check on our derivation. We first note, after applying equation~(47)
to the negative of equation~(39), that
\begin{equation}
{\dot J} \,=\, {{16\,\pi}\over 15}\,\,{q^2 \over (1\,+\,q)^4}\,\,
{\omega^3 \over c_s^5}\,\,\rho_0\,\,(G\,M_{\rm tot})^2\,\,a_{\rm tot}^4\,\,.
\end{equation}
If we then use equation~(44) to eliminate $G\,M_{\rm tot}$ in favor of 
$a_{\rm tot}$ and $\omega$, we obtain
\begin{displaymath}
{\dot J} \,=\, {{16\,\pi}\over 15}\,\,{q^2\over(1\,+\,q)^4}\,\,
{\omega^7 \over c_s^5}\,\,\rho_0\,\,a_{\rm tot}^{10} \,\,.
\end{displaymath}
Thus the energy emission rate can be written as
\begin{equation}
{\dot E} \,=\, {{16\,\pi}\over 15}\,\,{q^2\over(1\,+\,q)^4}\,\,
{\omega^8 \over c_s^5}\,\,\rho_0\,\,a_{\rm tot}^{10} \,\,.
\end{equation}
For fixed $a_{\rm tot}$, the binary components' relative speed scales with
$\omega$. This last expression thus reproduces the fact that the acoustic 
energy radiated by a quadrupolar source increases as the eighth power of the
Mach number \citep{L52}.  

\section{Binary Evolution}

We are now in a position to follow the binary's orbital decay in time. For this
purpose, we use for the torque the negative of $\dot J$ in equation~(51).
However, it is now appropriate to eliminate $a_{\rm tot}$, again employing
equation~(44):
\begin{equation}
\Gamma \,=\, -{{16\,\pi}\over 15}\,\,{q^2 \over (1\,+\,q)^4}\,\,
{\omega^{1/3} \over c_s^5}\,\,\rho_0\,\,(G\,M_{\rm tot})^{10/3}\,\,.
\end{equation}
We set this torque equal to the temporal derivative of $J_{\rm bin}$, as given
in equation~(48). Rearrangement gives an equation for the evolution of 
$\omega$: 
\begin{equation}
{\dot\omega} \,=\, {{16\,\pi}\over 5}\,{q\over (1\,+\,q)^2}\,
\left({G\,M_{\rm tot}}\over c_s^3\right)^{5/3}\!G\,\rho_0\,\,\omega^{5/3} 
\,\,.
\end{equation}

Equation~(54) is readily integrated. If $\omega_0$ is the initial,
nondimensional angular rotation rate, then 
\begin{equation}
\omega \,=\, \omega_0\,
\left(1\,-\,{t\over t_c}\right)^{-3/2} \,\,,
\end{equation}
where the coalesence time $t_c$ is
\begin{mathletters}
\begin{eqnarray}
t_c \,&\equiv&\, {{15}\over{32\,\pi}}\,{(1\,+\,q)^2 \over q}\,
{1\over{\rho_0\,G}}\,
\left({{G\,M_{\rm tot}}\over c_s^3}\right)^{-5/3}\!\!\omega_0^{-2/3} \\
&=&\, 2\times 10^5\,\,{\rm yr}\,
\left({n\over{10^7\,\,{\rm cm}^{-3}}}\right)^{-1}
\left({c_s \over{2\,\,{\rm km}\,\,{\rm s}^{-1}}}\right)^5
\left({M_{\rm tot}\over {1\,\,\Msun}}\right)^{-5/3}
\left({P_0\over{10^3\,\,{\rm yr}}}\right)^{2/3} \,\,.
\end{eqnarray}
\end{mathletters}
At time $t_c$,~$\omega$ diverges and the binary has contracted to zero 
separation. In our numerical evaluation of this time, we have set 
\hbox{$q\,=\,1$} and used the initial binary period $P_0$ in place of the 
angular velocity $\omega_0$. For \hbox{$M_{\rm tot}\,=\,\,1\,\,\Msun$}, a 
period of \hbox{$10^3\,\,{\rm yr}$} corresponds to 
\hbox{$a_{\rm tot} \,=\, 100\,\,{\rm AU}$}.\footnote{The evolutionary 
equation~(54) neglects accretion from the external medium. A numerical estimate
shows that the mass gain is not major for our adopted parameters, but it should
be included in a more complete analysis. According to equation~(56), any 
increase of $M_{\rm tot}$ shortens the coalescence time $t_c$.}

Finally, we may determine the mechanical energy release of the decaying binary 
as a function of time. Setting \hbox{${\dot E}\,=\,-\omega\,\Gamma$} and taking
$\Gamma$ from equation~(53), we have
\begin{mathletters}
\begin{eqnarray}
{\dot E} \,&=&\,{{16\,\pi}\over 15} \,{q^2\over(1\,+\,q)^4}\,
{\omega^{4/3}\over c_s^5}\,\rho_0\,(G\,M_{\rm tot})^{10/3} \\
&=&\,{\dot E}_0 \,\left(1\,-\,{t\over t_c}\right)^{-2} \,\,.
\end{eqnarray}
\end{mathletters}
Here, we have supplied the time dependence of $\omega$ from equation~(55). 
The constant ${\dot E}_0$ is
\begin{equation}
{\dot E}_0 \,\equiv\,{{16\,\pi}\over 15} \,{q^2\over(1\,+\,q)^4}\,\,
\omega_0^{4/3}\,\left({{G\,M_{\rm tot}}\over c_s^3}\right)^{10/3}\!\!
\rho_0\,c_s^5\,\,,
\end{equation}
and has the value $9\times 10^{-4}\,\,\Lsun$ for our fiducial parameters. 
Evidently, the energy release does not rival the radiative loss from the stars
themselves until very late during the inspiral, when $t$ is within a few
percent of $t_c$. At this point, the stars are separated by several AU.

\section{Discussion}

We may compare, at least in a qualitative manner, our derived torque with that
indicated by the traditional theory of dynamical friction. According to 
equation~(12) of \citet{o99}, the retarding force on a mass $M$ moving at speed
$V$ through a cloud of density $\rho_0$ is
\begin{displaymath}
F_{\rm DF} \,=\, -{{4\,\pi\,(G\,M)^2\,\rho_0}\over V^2}\,\,{\cal I} \,\,.
\end{displaymath}
The factor $\cal I$, essentially a Coulomb logarithm, is a nondimensional 
function of the Mach number $V/c_s$ and the time since the mass first entered
the cloud in question. Since $\cal I$ will generally be of order unity, we may
ignore it, along with other such factors, in the dimensional argument that
follows.

To apply this formula to the binary problem, we interpret $V$ as the 
components' relative velocity $V_{\rm rel}$, which is $\omega\,a_{\rm tot}$. 
Then the torque is of order \hbox{$a_{\rm tot}\,F_{\rm DF}$}, so that 
\begin{displaymath}
\Gamma_{\rm DF} \,\sim\, -{{\omega^3\,\rho_0\,G^2\,I^2}\over V_{\rm rel}^5} 
\,\,. 
\end{displaymath}
Here, the moment of inertia has been approximated as 
\hbox{$I\,\approx\,M\,a_{\rm tot}^2$}. Comparison to equation~(39) shows that
$\Gamma_{\rm DF}$ is the true $\Gamma$ multiplied by a factor 
\hbox{$(c_s/V_{\rm rel})^5$}. This factor can be much smaller than unity for 
the hard binaries of interest. On the other hand, the nondimensional $\cal I$ 
formally diverges for a velocity of $c_s$. The traditional theory is thus 
unreliable in this context.

However, it is not difficult to envision circumstances in which the present 
theory requires modification. From equation~(35a), the Mach number associated 
with the radial velocity amplitude is
\begin{eqnarray}
{u_r\over c_s} \,&=&\,{{G\,I\,\omega^2}\over{c_s^4\,r}} \\
                 &\sim&\,\left(V_{\rm rel}\over c_s\right)^2 
                 {r_{\rm in}\over r} \,\,.
\end{eqnarray}
Since \hbox{$r_{\rm in}\,\ll\,r$}, the induced velocity is normally subsonic.
For binaries that are initially very hard, or late during the inspiral of any
system, $u_r$ throughout the far field becomes supersonic. The disturbance then
changes character from an acoustic wave to a tightly wound spiral shock. In the
same regime, the density perturbation $\rho_1$ is comparable to or even exceeds
$\rho_0$, so that a fully nonlinear treatment is necessary. 

Additional modification of the theory would be required by the inclusion of a
finite eccentricity in the binary orbits. While the system would still be
periodic, the equivalent quadrupolar source would exhibit variation over a 
continuous distribution of frequencies. The same frequency distribution would 
then appear in the transmitted waves. It would be interesting to recalculate 
both the torque and energy loss under this more general condition and thereby 
follow the evolution of the eccentricity during orbital decay.

Returning to the theory's main astrophysical application, our result for the
coalescence time is a reasonable one that adds credence to the underlying
picture. The nearest and best-studied region of massive star formation is the
Orion Nebula Cluster, whose general population formed 1-2~Myr ago
\citep{h97}. We do not know the age of the high-mass members, the Trapezium,
with any precision. \citet{ps01} have argued that they are relatively young,
of order $10^5~{\rm yr}$, based on the location of BM~Ori and its binary
companion in the HR diagram. Assuming that the Trapezium stars, along with 
their companions, are coeval, their inferred age provides an upper bound to the
formation time scale. We are thus encouraged by this matching of times. We
stress, however, that our expression for $t_c$ in equation~(56) varies 
inversely with the imprecisely known ambient density $n$. 

If massive stars indeed coalesce over such a period, perhaps they do so through
an accelerating sequence of binary mergers. The binaries themselves might be 
created and disrupted rapidly out of the dense gas, leading to a statistically 
stable period distribution, as originally envisioned by \citet{lb69}. Within 
this picture, one could in principle determine the mass distribution of the 
growing population of coalesced objects, thereby advancing the theory another
significant step.

\acknowledgments

This project was originally inspired by extensive discussions with Avery 
Broderick concerning the formation of massive stars. Kevin Bundy provided 
useful comments on a preliminary draft of the manuscript. The author was 
partially supported by NSF grant AST-0908573. 

\clearpage

\appendix

\section{Gravitational Potential from the Equivalent Density}

We wish to verify that $\rho_1^\ast$ in equation~(22) indeed generates the
gravitational potential $\Phi_1^\ast$ in equation~(16). It is evident that the
monopolar portion $\rho_{10}^\ast$ in equation~(17) does yield the
term $G\,M_{\rm tot}/r$. We therefore focus on $\rho_{12}^\ast$, given in 
equations~(20)-(21), and verify that
\begin{equation}
\Phi_{12}^\ast (\bx) \,=\, -G\int\!
{\rho_{12}^\ast (\bx^\prime)\over\vert\bx-\bx^\prime\vert}  
\,\,d^3 x^\prime\,\,,
\end{equation}
where
\begin{equation}
\Phi_{12}^\ast \,\equiv\, -{{G\,I}\over{2\,r^3}}
\left[{3\over 2}\,{\rm sin}^2\,\theta\,\,
{\rm cos}\,2 (\omega t \,-\,\phi) \,-\,
\left( {3\over 2} {\rm cos}^2\,\theta \,-\,{1\over 2} \right)\right] \,\,.
\end{equation}

The derivation proceeds in a manner analogous to the calculation of $\rho_1$ in
Section~4.2. We first find three functions ${\cal F}_A$, ${\cal F}_B$, and 
${\cal F}_C$ obeying    
\begin{mathletters}
\begin{eqnarray}
\nabla^2{\cal F}_A \,&=&\, 2\,\pi\,G\,I\,\,{\rm cos}^2 \omega t\,\,\delta (x)
\,\delta (y)\,\delta (z)
\\
\nabla^2{\cal F}_B \,&=&\, 2\,\pi\,G\,I\,\,{\rm sin}^2 \omega t\,\,\delta (x)
\,\delta (y)\,\delta (z)
\\
\nabla^2{\cal F}_C \,&=&\, 2\,\pi\,G\,I\,\,{\rm sin}\,2\,\omega t\,\,\delta (x)
\,\delta (y)\,\delta (z) \,\,.
\end{eqnarray}
\end{mathletters}
If we further define
\begin{mathletters}
\begin{eqnarray}
\Phi_A^\ast \,&\equiv&\, {{\partial^2 {\cal F}_A}\over
{\partial x^2}}
\\
\Phi_B^\ast \,&\equiv&\, {{\partial^2 {\cal F}_B}\over
{\partial y^2}}
\\
\Phi_C^\ast \,&\equiv&\, {{\partial^2 {\cal F}_C}\over
{\partial x\,\partial y}}
\,\,,
\end{eqnarray}
\end{mathletters}
then the combination
\begin{equation}
\Phi_{12}^\ast \,=\, \Phi_A^\ast \,+\, \Phi_B^\ast \,+\, \Phi_C^\ast \,\,,
\end{equation}
obeys Poisson's equation with $\rho_{12}^\ast$ as the source density. 
Equivalently, $\Phi_{12}^\ast$ is the solution of equation~(A1).

The functions ${\cal F}_A$, ${\cal F}_B$ and ${\cal F}_C$ are all solutions
of Poisson's equation with a central point mass:
\begin{mathletters}
\begin{eqnarray}
{\cal F}_A \,&=&\, -{{G\,I}\over{2\,r}}\,\,{\rm cos}^2\,\omega t \\
{\cal F}_B \,&=&\, -{{G\,I}\over{2\,r}}\,\,{\rm sin}^2\,\omega t \\
{\cal F}_C \,&=&\, -{{G\,I}\over{2\,r}}\,\,{\rm sin}\,2\,\omega t \,\,.
\end{eqnarray}
\end{mathletters}
By successive differentiation of $1/r$, we find
\begin{mathletters}
\begin{eqnarray}
\Phi_A^\ast \,&=&\, -{{G\,I}\over 2}\,
\left(-{1\over r^3} \,+\, {{3 x^2}\over r^5}\right) 
{\rm cos}^2\,\omega t  \\
\Phi_B^\ast \,&=&\, -{{G\,I}\over 2}\,
\left(-{1\over r^3} \,+\, {{3 y^2}\over r^5}\right) 
{\rm sin}^2\,\omega t  \\
\Phi_C^\ast \,&=&\, -{{G\,I}\over 2}\,
{{3 x y}\over r^5}\,
{\rm sin}\,2\,\omega t \,\,.
\end{eqnarray}
\end{mathletters}
Adding these components yields
\begin{eqnarray}
\Phi_{12}^\ast \,&=&\,-{{G\,I}\over{2\,r^3}} \,\,[ 
{\rm cos}^2\,\omega t\, 
\left( -1 \,+\, 3\,{\rm sin}^2\,\theta\,\,{\rm cos}^2\,\phi\right)\nonumber\\ 
&{\phantom =}& \ \ \ \ \ \ \  +\,\,{\rm sin}^2\,\omega t\,  
\left( -1 \,+\, 3\,{\rm sin}^2\,\theta\,\,{\rm sin}^2\,\phi\right)\nonumber \\
&{\phantom =}& \ \ \ \ \ \ \  +\,\,{\rm sin}\,2\,\omega t\,
\left(3\,{\rm sin}^2\theta\,\,{\rm sin}\,\phi\,\,{\rm cos}\,\phi\right)
] \,\,,
\end{eqnarray}
which simplifies to equation~(A2).

\clearpage

\begin{figure}
\plotone{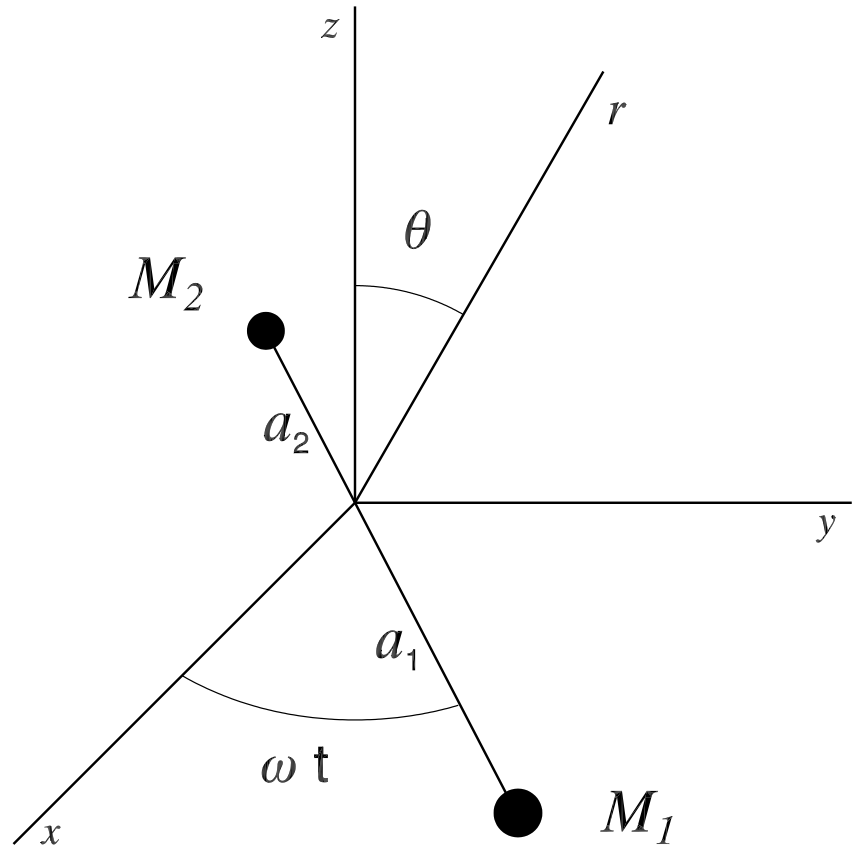}
\caption{Basic binary parameters. Two stars, $M_1$ and $M_2$, orbit their
common center of mass, which is at the origin of the coordinate system. The
circular orbits, of radii $a_1$ and $a_2$, respectively, lie in the 
$x$-$y$ plane. At the time shown, $M_1$ has swept out the angle
\hbox{$\phi^\prime\,=\,\omega\,t$} from the $x$-axis.}
\end{figure}

\begin{figure}
\plotone{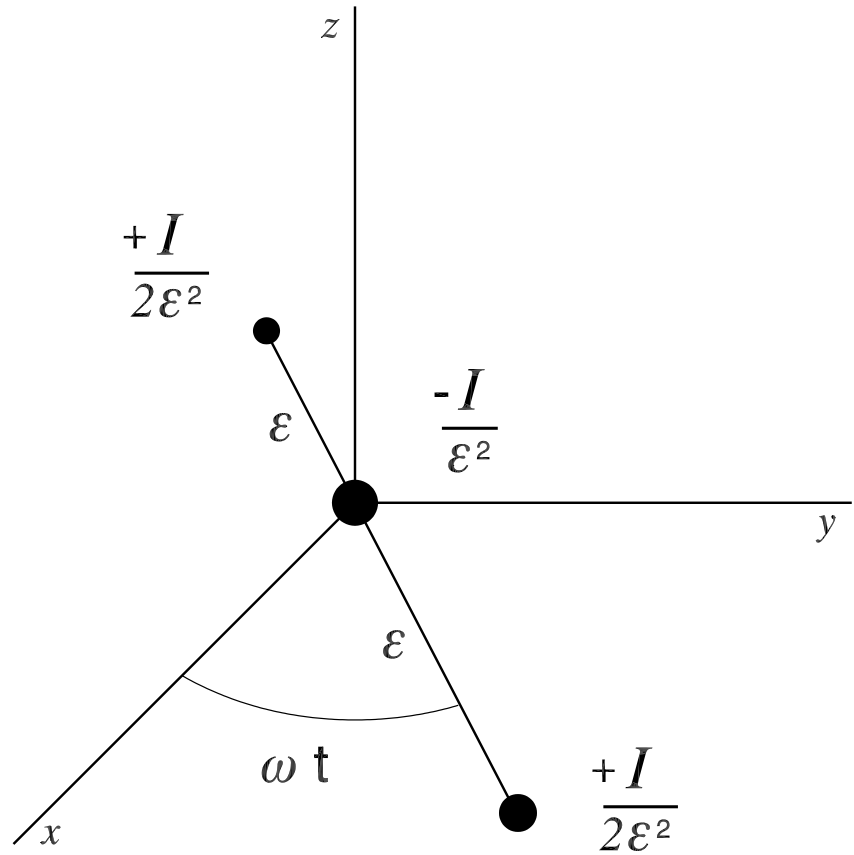}
\caption{Equivalent system of masses that gives the same gravitational
potential, through quadrupolar order, as the actual binary. Two equal
masses lie at a distance $\epsilon$ on either side of the rotation axis, 
while a negative mass of twice the magnitude lies at the center.}
\end{figure}

\begin{figure}
\plotone{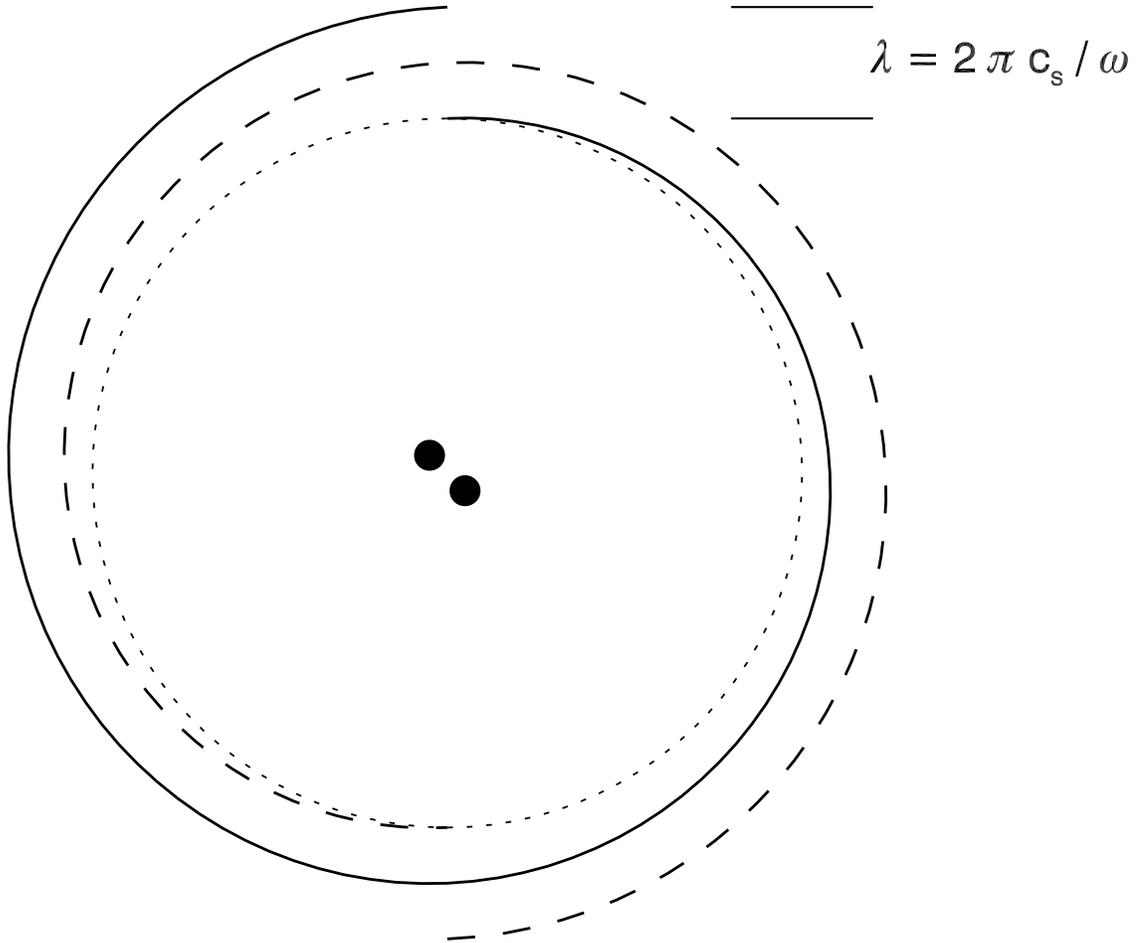}
\caption{Geometry of the acoustic wave. The radial wavelength of a single 
spiral arm is $\lambda$, as shown. However, because a second arm is 
interleaved, the actual distance between successive wave crests is 
$\lambda/2$. The central binary is rotating counterclockwise at angular
speed $\omega$.}
\end{figure}

\end{document}